\documentclass[12pt,indent]{article}
\usepackage{a4,latexsym,amsmath,amssymb}
\usepackage[latin1]{inputenc}
\usepackage{float}
\usepackage{natbib}
\usepackage{graphics}
\usepackage[english]{babel}
\usepackage{tabularx}
\usepackage{lscape}
\usepackage{multirow}
\usepackage{rotating}
\usepackage{dsfont}
\usepackage{caption}
\usepackage{subcaption}
\usepackage{bbm}
\usepackage[table]{xcolor}
\usepackage{booktabs}
\usepackage{calc}
\usepackage{ifthen}
\usepackage{url}
\usepackage{colortbl}      
\usepackage{tikz}

\usepackage{blindtext}
\usepackage{scrextend}
\usepackage{mathtools}
\usepackage{verbatim}

\usetikzlibrary{shapes.geometric, chains, calc}

\newenvironment{tabularsmall}
{ \footnotesize \sffamily \tabular } {
\endtabular
\normalfont }

\newcommand{\logit}{\operatorname{logit}}
\newcommand{\e}{\operatorname{e}}

\newcommand{\alphab}{\boldsymbol{\alpha}}

\newcommand{\betab}{{\boldsymbol{\beta}}}

\newcommand{\gammab}{\boldsymbol{\gamma}}

\newcommand{\cb}{\boldsymbol{c}}

\newcommand{\xb}{\boldsymbol{x}}

\newcommand{\zb}{\boldsymbol{z}}

\newcommand{\blanco}[1]{}

\def\d{\displaystyle}

\usepackage{natbib}

\usepackage{geometry}
\usepackage{setspace}
\usepackage{tikz}
\usepackage[]{graphicx}
\usepackage[]{color}
\makeatletter
\def\maxwidth{ %
  \ifdim\Gin@nat@width>\linewidth
    \linewidth
  \else
    \Gin@nat@width
  \fi
}
\makeatother

\definecolor{fgcolor}{rgb}{0.345, 0.345, 0.345}

\usepackage{framed}
\makeatletter
 {\par\unskip\endMakeFramed%
 \at@end@of@kframe}
\makeatother

\definecolor{shadecolor}{rgb}{.97, .97, .97}
\definecolor{messagecolor}{rgb}{0, 0, 0}
\definecolor{warningcolor}{rgb}{1, 0, 1}
\definecolor{errorcolor}{rgb}{1, 0, 0}

\usepackage{alltt}
\IfFileExists{upquote.sty}{\usepackage{upquote}}{}

\begin{document}
\bibliographystyle{chicago}
\sloppy

\makeatletter
\renewcommand{\section}{\@startsection{section}{1}{\z@}%
        {-3.5ex \@plus -1ex \@minus -.2ex}%
        {1.5ex \@plus.2ex}%
        {\reset@font\large\sffamily}}
\renewcommand{\subsection}{\@startsection{subsection}{1}{\z@}%
        {-3.25ex \@plus -1ex \@minus -.2ex}%
        {1.1ex \@plus.2ex}%
        {\reset@font\normalsize\sffamily\flushleft}}
\renewcommand{\subsubsection}{\@startsection{subsubsection}{1}{\z@}%
        {-3.25ex \@plus -1ex \@minus -.2ex}%
        {1.1ex \@plus.2ex}%
        {\reset@font\normalsize\sffamily\flushleft}}
\makeatother



\newsavebox{\tempbox}
\newlength{\linelength}
\setlength{\linelength}{\linewidth-10mm} \makeatletter
\renewcommand{\@makecaption}[2]
{
  \renewcommand{\baselinestretch}{1.1} \normalsize\small
  \vspace{5mm}
  \sbox{\tempbox}{#1: #2}
  \ifthenelse{\lengthtest{\wd\tempbox>\linelength}}
  {\noindent\hspace*{4mm}\parbox{\linewidth-10mm}{\sc#1: \sl#2\par}}
  {\begin{center}\sc#1: \sl#2\par\end{center}}
}



\def\R{\mathchoice{ \hbox{${\rm I}\!{\rm R}$} }
                   { \hbox{${\rm I}\!{\rm R}$} }
                   { \hbox{$ \scriptstyle  {\rm I}\!{\rm R}$} }
                   { \hbox{$ \scriptscriptstyle  {\rm I}\!{\rm R}$} }  }

\def\N{\mathchoice{ \hbox{${\rm I}\!{\rm N}$} }
                   { \hbox{${\rm I}\!{\rm N}$} }
                   { \hbox{$ \scriptstyle  {\rm I}\!{\rm N}$} }
                   { \hbox{$ \scriptscriptstyle  {\rm I}\!{\rm N}$} }  }

\def\d{\displaystyle}\def\d{\displaystyle}

\title{Non Proportional Odds Models are  Widely Dispensable  - 
 Sparser Modeling based on  Parametric and Additive Location-Shift Approaches
}
  \author{Gerhard Tutz$^*$    \& Moritz Berger$^{**}$ \vspace{0.4cm}\\
{\small $^*$Ludwig-Maximilians-Universit\"{a}t M\"{u}nchen}\\
{\small $^{**}$ Institut für Medizinische Biometrie, Informatik und Epidemiologie, }\\ {\small  Medizinische Fakultät, Universität Bonn}}

\maketitle

\begin{abstract} 
\noindent
The potential of location-shift models to find adequate models between the proportional odds model and the non-proportional odds model is investigated. It is demonstrated that these models are very useful in ordinal modeling. While   proportional odds models are often too simple, non proportional odds models are typically unnecessary complicated and seem widely dispensable. The class of location-shift models is also extended to allow for smooth effects. The additive location-shift model contains two functions for each explanatory variable, one for the location and one for dispersion. It is much sparser than hard-to-handle additive models with category-specific covariate functions but more flexible than common vector generalized additive models.
\end{abstract}


\section{Introduction}
The proportional odds model, which was propagated by \citet{McCullagh:80} is probably the most widely used ordinal regression model.  
The assumption that effects of covariates are not category-specific makes it a simply structured  model that allows to interpret   parameters  in terms of cumulative odds. However, in many applications the model  shows poor goodness-of-fit and does not   adequately represent the underlying probability structure.
As alternatives non proportional and partial proportional odds models were proposed. They allow for category-specific effects of explanatory variables, and typically   show much better fit than proportional odds models see, for example, \citet{Brant:90}, \citet{PetHar:90},   \citet{BenGro:98}, \citet{Cox:95}, \citet{kim2003assessing}, \citet{williams2006generalized},  \citet{liu2009graphical}, and  \citet{williams2016understanding}.

A major disadvantage of non proportional odds models is that many parameters are involved, which makes interpretation of parameters much harder than in the simple proportional odds model. Moreover, the space of explanatory variables can be almost empty, and estimation of parameters tends to fail in cases with a larger number of response categories. In the present paper models between the proportional and non proportional odds models are propagated. They are sufficiently complex to provide an adequate fit but contain much less parameters than the non proportional odds model.

Non proportional and proportional   odds model are logistic versions of cumulative ordinal models with  category-specific or global,  that is, not category-specific  effects of variables, respectively. We consider the more general class of cumulative models, which may use any response function that is determined by a strictly increasing distribution function. In addition, we consider the alternative class of adjacent categories models with general link functions. For all of these models it is essential to find an adequate representation of data that does not involve too many parameters.

The class of models that is investigated contains a location term in the tradition of the proportional odds model (and other models with global parameters), but instead of using a multitude of category-specific parameters, the location term is complemented by a linear term that represents variability of the response, which may be seen as  dispersion or, in questionnaires, the tendency of respondents to prefer extreme or middle categories. Parametric models of this type 
were considered by \citet{TuBerg2016RespStyle, TuBer2017Disp}.

The paper has two main objectives. It is demonstrated that location-shift versions of cumulative and adjacent categories models are often adequate when modeling ordinal responses. They can be seen as a natural extension of proportional odds models to complexer models that avoid the complexity of non proportional odds models. Indeed, non proportional odds type models turn out to be a frequently dispensable class of models. They are unnecessarily complicated, and hardly needed in ordinal modelling. 
In contrast to most statistical papers, which propagate  more complex modeling, in the first part of the paper  we plead for a simpler class of models instead of a complexer one. In the second part of the paper  location-shift models are extended to allow for smooth effects of covariates. Extensions of additive models that include general category-specific effects are rather hard  to obtain. The proposed additive location-shift model offers a way to go beyond the simple global effects model without adding too many functions.

The main messages of the paper can be summarizes as follows
\begin{itemize}
\item Proportional  odds models, or, more general, models with category-specific parameters are widely dispensable. In many applications a simpler version is appropriate.
\item Location-shift models, which  are propagated here, have the advantage that they show not only location effects but dispersion effects or tendencies to respond,  which are typically present in applications.
\item If linear effects are questionable, the smooth location-shift model provides an alternative to simple global effect models. The models allow to account for smooth dispersion effects. 
\end{itemize}

In Section \ref{sec:ordinal} parametric ordinal models and their location-shift versions are considered. It is demonstrated that in applications non proportional odds models are often not needed. In Section \ref{sec:additive} traditional additive ordinal models are briefly considered and the additive location-shift model  is introduced as an alternative.

\section{Ordinal Regression Models }\label{sec:ordinal}

\subsection{Proportional and Non Proportional Odds Models }
The most widely used ordinal regression model is the proportional odds model, which is a member of the class of cumulative models. {Cumulative models} can be derived from an underlying latent variable. Let $Y^*$ be an underlying latent variable for which the regression model $Y^*=-\xb^T\betab+\epsilon$ holds, where $\epsilon$ is a noise variable with continuous distribution function $F(.)$, $\xb$ is a vector of explanatory variables, and  $\betab$ a vector of coefficients. If one assumes that the link between the observable categorical response $Y$ and the latent trait is specified by  $Y=r \Leftrightarrow \theta_{r-1}< Y^*\leq\theta_{r}$, where $-\infty =\theta_{0}<\theta_{1}<\dots<\theta_{k}=\infty$, one obtains the \textit{cumulative model}
\begin{equation}\label{eq:cum}
P(Y \le r|\xb )= F(\beta_{0r}+\xb^T\betab), \quad r=1,\dots,k-1,
\end{equation}
where the category-specific intercepts $\beta_{0r}$ are identical to the thresholds on the latent scale, that is, $\beta_{0r}=\theta_{r}$. If one uses the logistic distribution $F(\eta)=\exp(\eta)/(1+\exp(\eta))$, one obtains the \textit{proportional odds model}
\[
\logit P(Y \le r|\xb) = \beta_{0r}+\xb^T\betab.
\]
The strength of the model is that interpretation of parameters is very simple. Let  $\gamma_r(\xb)= {P(Y \le r|\xb)}/{P(Y>r|\xb)}$ denote the cumulative odds, then 
$\e^{\beta_{j}}$ can be  directly interpreted as the  odds ratio that compares the cumulative odds with value $x_j+1$ in  the $j$-th variable  to the  odds with value $x_j$ in   the $j$-th variable, when all other variables are kept fixed, 
\begin{equation}\label{eq:int}
\e^{\beta_{j}}=\frac{\gamma_r(x_1,\dots, x_j+1,\dots,x_p)}{\gamma_r(x_1,\dots, x_j,\dots,x_p)}.
\end{equation}
It is important that the interpretation does not depend on the category, $\e^{\beta_{j}}$ is the same for all odds $\gamma_r$, $r=1,\dots,k-1$. The independence of parameters on categories holds for the whole class of cumulative models (\ref{eq:cum}) since they share the stochastic ordering property, which means that for two sets of explanatory variables $\xb$ and $\tilde\xb$ the term 
\[
F^{-1}(P(Y\leq r|\xb))-F^{-1}(P(Y\leq r|\tilde\xb))=(\xb-\tilde\xb)^T\gammab,
\]
does not depend on the category $r$.

Early versions of the cumulative logistic model were given by
\citet{Snell:64}, \citet{WalDun:67}, and \citet{WilGri:72}. More general cumulative models were considered, among others, by \citet{ArmSlo:89},   \citet{GenFar:85}, \citet{AnaKlei:97}, \citet{SteWei:98}. \citet{Rud-etal:95} and \citet{CamDon:89} investigated their use in prediction, more recently, robust estimators have been proposed   by \citet{iannario2017robust}.

The problem with cumulative models of the form (\ref{eq:cum}) is that they often do  not fit the data well, which calls for more complicated models. A class of models that has been considered in the literature is the \textit{cumulative model with category-specific effects}
\begin{equation}\label{eq:cum2}
P(Y \le r|\xb )= F(\beta_{0r}+\xb^T\betab_r), \quad r=1,\dots,k-1,
\end{equation}
which uses the parameter vectors $\betab_r^T=(\beta_{1r},\dots,\beta_{pr})$, and allows that parameters vary across categories. Logistic models of this type are also called \textit{non proportional odds models} to distinguish them from the simpler versions. If $\betab_1=\dots=\betab_{k-1}$, the model simplifies to the simple cumulative model (\ref{eq:cum}). Of course the parameters have not to vary over categories for all variables, which might yield two types of variables, variables with a \textit{global} effect, for which $\beta_{j1}=\dots=\beta_{j,k-1}=\beta_{j}$, and variables with   \textit{category specific effects}, that is $\beta_{js}\ne \beta_{jr}$ for at least two categories $s,r$. 
Logistic versions of this   type of model are called \textit{partial proportional odds models} and have been investigated, for example, by \citet{Brant:90}, \citet{PetHar:90},   \citet{BenGro:98}, \citet{Cox:95}, \citet{kim2003assessing} and \citet{liu2009graphical}.
In sociology they are also referred to as generalized ordered logit models \citep{williams2006generalized,williams2016understanding}.

The general model with category-specific effects is attractive since it usually provides a better fit to the data. However, the whole class of models has some serious disadvantages. One is that one has many parameters, which are   much harder to interpret. More seriously,  the possible values of explanatory variables can be strongly restricted because it is postulated that  $\beta_{01} + \xb^T\betab_1 \le \dots \le \beta_{0,k-1} + \xb^T\betab_{k-1}$ for all values $\xb$. Even if estimates exist, in future observations with more extreme values in the explanatory variables  the estimated probabilities can be negative. For problems with the model see also \citet{walker2016generalizing} who even concludes   that it is impossible to generalize the cumulative class of ordered regression models in ways consistent with the
spirit of generalized cumulative regression models.

\subsection{Cumulative Location-Shift Models }

An alternative extension of the non proportional odds model was proposed  by \citet{TuBer2017Disp}. They  assume that variables may change the thresholds of the underlying latent trait. Then, thresholds  $\beta_{0r}$ in the proportional odds model are replaced by $\beta_{0r}+ (k/2 -r) \zb^T\alphab$, where  $\zb$ denotes a vector of covariates, possibly containing components of $\xb$. The replacement yields the  so-called location-shift model, which is given in closed form by  
\begin{equation}\label{eq:locsh}
P(Y \le r|\xb )= F(\beta_{0r}+\xb^T\betab+ (r -k/2) \zb^T\alphab), \quad r=1,\dots,k-1.
\end{equation}
It contains the familiar location term $\xb^T\betab$, which models the location on the latent continuum and therefore the tendency to low or high response categories. In addition it contains the scaled shifting term $(r -k/2) \zb^T\alphab$, which modifies the thresholds and has a quite different interpretation.

The term $\zb^T\alphab$ determines the shifting of thresholds, whereas the scaling factor $(r-k/2)$ is an additional weight chosen such that the difference between thresholds are widened or shrunk by the same amount. 
For illustration let us consider the case $k=6$, for which  the modified thresholds  $\beta_{0r}+(r -k/2) \zb^T\alphab$ have the form
\begin{figure}[H]
\centering
\begin{tikzpicture}[scale=1]
    \tikzstyle{ann} = [draw=none,fill=none]
    \matrix[nodes={draw,   thick},
        row sep=0.3cm,column sep=0.82cm,
				column 1/.style={nodes={rectangle, draw, minimum width=2em}},
				column 3/.style={nodes={rectangle, draw, minimum width=2em}},
				column 5/.style={nodes={rectangle, draw, minimum width=2em}},
				column 7/.style={nodes={rectangle, draw, minimum width=2em}},
				column 9/.style={nodes={rectangle, draw, minimum width=2em}},
				column 11/.style={nodes={rectangle, draw, minimum width=2em}}
				] {
				    \node[circle] {1}; &\draw [line width=0.03cm] (4,0.7) -- (4,-0.7); &			    \node[circle] {2};&\draw [line width=0.03cm] (4,0.7) -- (4,-0.7); &			    \node[circle] {3};&\draw [line width=0.03cm] (4,0.7) -- (4,-0.7); &				    \node[circle] {4};&\draw [line width=0.03cm] (4,0.7) -- (4,-0.7); &	    \node[circle] {5};&\draw [line width=0.03cm] (4,0.7) -- (4,-0.7); &	    \node[circle] {6};\\
				};
\end{tikzpicture}
\begin{tikzpicture}[scale=1]
    \tikzstyle{ann} = [draw=none,fill=none]
    \matrix[nodes={draw,   thick}, 
        row sep=0.3cm,column sep=2.2cm,
								column 1/.style={nodes={rectangle, draw, minimum width=2cm}}
				] {
				   \hspace{1cm} &\small $\beta_{01}-2\zb^T\alphab$; 	 &   &\small $\beta_{02}-\zb^T\alphab$; 	&	&\phantom{$123$}$\beta_{03}$	 \hspace{0.8cm}   & 				  &\small $\beta_{04}+\zb^T\alphab$\phantom{$ $};&				  &\small $\beta_{05}+ 2\zb^T\alphab$;   \\
				};
\end{tikzpicture}
\end{figure}
In general, thresholds are widened if $\zb^T\alphab$ is positive, and shrunk if it is negative. The consequence is that one observes more concentration in the middle or extreme  categories, respectively. Since more concentration in the middle means less variability the term $\zb^T\alphab$ can also be seen as representing dispersion. For positive values the distribution is more concentrated, meaning smaller dispersion, for negative values one has larger dispersion. The effect is also seen from considering the differences between adjacent predictors,
\begin{equation}\label{eq:eta}
\eta_r -\eta_{r-1}=\beta_{0r}-\beta_{0,r-1}+ \zb^T\alphab, \quad r=2,\dots,k-1,
\end{equation}
where $\eta_r =\beta_{0r}+\xb^T\betab+ (r -k/2) \zb^T\alphab$ is the $r$-th predictor. 
For positive values of $\zb^T\alphab$ the difference between adjacent predictors becomes greater, for negative values it becomes smaller. Thus, $\alphab$ represents the tendency to  middle or extreme  categories linked to covariates $\zb$, which is separated from the  location effect   $\xb^T\betab$. It can be derived that the interpretation of the $\beta$ parameters is the same as in the proportional odds model if $\xb$ and $\zb$ are distinct \citep{TuBer2017Disp}.

A different view of the model is obtained by seeing it as a non proportional odds model with specific constraints on the parameters. 
Let us consider the general case $\xb=\zb$. Then one has 
\[
P(Y \le r|\xb )= F(\beta_{0r}+\xb^T(\betab+ (r -k/2) \alphab)) = F(\beta_{0r}+\xb^T\betab_r),
\]
where $\betab_r=\betab+ (r -k/2) \alphab$. The model is equivalent to a  category-specific model with  constraints 
\[
(\betab_r- \betab)/ (r -k/2)= \cb, \quad r=1,\dots,k-1,
\]
where $\betab=\sum_{r=1}^{k-1}\betab_r$, and $\cb$ is a vector of constants. If the category-specific model with  constraints is assumed to hold, the vector $\cb$ turns out to be $\alphab$.

That means, in particular, that the location-shift model is a submodel of the model with category-specific effects. Since the proportional odds model is a submodel of the location-shift model one has the nested structure 
\[
\text{proportional odds model} \subset \text{location-shift  model}\subset \text{non proportional odds model}
\]  
or, more generally,
\[
\text{model with global effects} \subset \text{location-shift  model}\subset \text{model with category-specific effects}.
\]
Since the location-shift  model is   a (multivariate) generalized linear model one can investigate if the  models can be simplified by testing the sequence of nested models, see, for example, \citet{TutzBook2011}.

One of the disadvantages of model versions with category-specific effects is that the simple interpretation of parameters gets lost. One has a multitude of parameters for which one might easily lose track. For example, if one has just four variables and ten categories (see the example in Section \ref{sec:ex}) the model contains 45 parameters, for each variable one has 9 parameters. In contrast, the proportional odds model contains only 13 parameters, the impact of one variable is described by just one parameter.
The models propagated here  are models that are between the most general model and the model with global effects,  in them the impact of a single variable is described by just two parameters (instead of $k-1$ parameters as in the general model and one in the model with global effects).

\subsection{An Example: Safety in Naples}\label{sec:ex}
The package CUB \citep{iannario2018cub} contains the data set relgoods, which provides results of a survey aimed at measuring the subjective extent of feeling safe in the streets. The data were collected  in the metropolitan area of Naples, Italy. Every participant was asked to assess on a 10 point ordinal scale his/her personal score for  feeling  safe with large categories referring to feeling  safe.
There are $n=2225$ observations and four variables,  \textit{Age}, \textit{Gender} (0: male, 1: female), \textit{Residence} (1: City of Naples, 2: District of Naples, 3: Others Campania, 4: Others Italia) and the educational degree (\textit{EduDegree}; 1: compulsory school, 2: high school diploma, 3: Graduated-Bachelor degree, 4: Graduated-Master degree, 5: Post graduated).

\begin{table}[!ht]
\caption{Fits of cumulative models with logistic link for safety data.}\label{tab:nuccatspec1} 
\centering
\begin{tabularsmall}{lrrrrrrrrrr}
  \toprule
 & deviance     & df  &  difference in   & df &$p$-value\\ 
 &      &   &deviances     & \\
  \midrule

Non proportional odds model               &9825.78     &19935 \\
Location-shift model                      &9899.67     &19998  & 73.89  &63 &0.1640\\
Proportional odds model                   &9948.99    &20007  &49.32  &9 &0.0000\\ 
\bottomrule  
\end{tabularsmall}
\end{table}

Table \ref{tab:nuccatspec1} shows the deviances of the fitted models and the differences. The full model with category-specific effects has 90 parameters, which reduces to 27 parameters in the location-shift model. The difference in deviances suggests that the full model can  be simplified to the location-shift model, however, it certainly does not simplify to the model with global effects (difference of deviances 49.32 on 9 df). That means the location-shift model contains enough structure to explain the effect of covariates on the response, but the simpler structure without the term $(r-k/2) \zb^T\alphab$ is too simple, that is, relevant effects are missing. It is noteworthy that in this application, as in the other applications used here, the sample size is rather large ($n=2225$). Typically, if sample sizes are large one finds more significant effects. Therefore, it is remarkable that the complex model with category-specific effects can be simplified in spite of the large sample size.

\subsection{Adjacent Categories Models }
An alternative class of models for ordinal responses  are  \textit{adjacent categories models}. In its simple  version  they assume
\begin{equation}\label{eq:adj}
P(Y > r|Y \in \{ r ,r+1\}, \xb)= F(\beta_{0r}+\xb^T\betab ), \quad r=1,\dots,k-1.
\end{equation}
where $F(.)$ again is a strictly increasing distribution function but no ordering of intercepts has to be postulated. The logistic version has the form
\begin{equation}\label{eq:adj2}
\log \left(\frac{P(Y = r+1|\xb )}{P(Y = r|\xb )}\right)= \beta_{0r}+\xb^T\betab , \quad r=2,\dots,k-1.
\end{equation}
The interpretation of parameters is as simple as for basic cumulative models; $\e^{\beta_{j}}$ is the  odds ratio that compares the  odds with value $x_j+1$ in  the $j$-th variable  to the  odds with value $x_j$ in   the $j$-th variable, however the odds are not cumulative odds but adjacent categories odds,  
$\gamma_r(\xb)=  {P(Y=r+1|\xb)}/{P(Y=r|\xb)}$.

In the same way as in the cumulative models the linear predictor can be replaced by a predictor  with category-specific parameters, that is, predictors $\eta_r=\beta_{0r}+\xb^T\betab_r$ to obtain a better fit. The   corresponding model contains   many parameters,  which are harder to interpret. A sparser model is the \textit{location-shift version of the adjacent categories model} 
\begin{equation}\label{eq:adj3}
\log \left(\frac{P(Y = r+1|\xb )}{P(Y = r|\xb )}\right)= \beta_{0r}+\xb^T\betab + (k/2 -r) \zb^T\alphab, \quad r=1,\dots,k-1.
\end{equation}

For $k=6$ one obtains for the term $(k/2 -r) \zb^T\alphab$, which distinguishes between category $r$ and $r+1$
\begin{figure}[H]
\centering
\begin{tikzpicture}[scale=1]
    \tikzstyle{ann} = [draw=none,fill=none]
    \matrix[nodes={draw,   thick},
        row sep=0.3cm,column sep=0.82cm,
				column 1/.style={nodes={rectangle, draw, minimum width=2em}},
				column 3/.style={nodes={rectangle, draw, minimum width=2em}},
				column 5/.style={nodes={rectangle, draw, minimum width=2em}},
				column 7/.style={nodes={rectangle, draw, minimum width=2em}},
				column 9/.style={nodes={rectangle, draw, minimum width=2em}},
				column 11/.style={nodes={rectangle, draw, minimum width=2em}}
				] {
				    \node[circle] {1}; &\draw [line width=0.03cm] (4,0.7) -- (4,-0.7); &			    \node[circle] {2};&\draw [line width=0.03cm] (4,0.7) -- (4,-0.7); &			    \node[circle] {3};&\draw [line width=0.03cm] (4,0.7) -- (4,-0.7); &				    \node[circle] {4};&\draw [line width=0.03cm] (4,0.7) -- (4,-0.7); &	    \node[circle] {5};&\draw [line width=0.03cm] (4,0.7) -- (4,-0.7); &	    \node[circle] {6};\\
				};
\end{tikzpicture}
\begin{tikzpicture}[scale=1]
    \tikzstyle{ann} = [draw=none,fill=none]
    \matrix[nodes={draw,   thick}, 
        row sep=0.3cm,column sep=2.2cm,
								column 1/.style={nodes={rectangle, draw, minimum width=2cm}}
				] {
				   \hspace{1cm} &\small $2\zb^T\alphab$; 	 &   &\small $\zb^T\alphab$; 		&	 \hspace{1.3cm}   &\small $0$; 				  &\small $-\zb^T\alphab$\phantom{$2$};&				  &\small $- 2\zb^T\alphab$;   \\
				};
\end{tikzpicture}
\end{figure}
Thus, one obtains the same effects as in the cumulative location-shift model, if $\zb^T\alphab$ is large the person has a tendency to choose middle categories, if $\zb^T\alphab$ is small there is a tendency to extreme categories. 

\blanco{
It is noteworthy that scaling is chosen such that in both models the cumulative and the adjacent categories model large values of $\zb^T\alphab$ correspond to a tendency to middle categories. Therefore the predictor contains the term $(r-k/2) \zb^T\alphab$ in the cumulative model but  $(k/2 -r) \zb^T\alphab= -(r-k/2) \zb^T\alphab$ in the adjacent categories model. 
It is important that interpretation changes if one uses different representations of the model. Advanced program packages allow to use reverse categories. Then, the cumulative location-shift model has the form   
\[
P(Y \ge r|\xb )= F(\beta_{0r}+\xb^T\betab+ (r -k/2) \zb^T\alphab), \quad r=2,\dots,k.
\]
However, then the effect of the shifting term is also reversed,   large values $\zb^T\alphab$ indicate a tendency to extreme categories while small values $\zb^T\alphab$ indicate a tendency to middle categories. Of course, also the meaning of the location effect $\xb^T\betab$ has changed, large values $\zb^T\alphab$ indicate a tendency to larger categories.
}

For the adjacent categories model the same hierarchy holds as for the cumulative models. The model with global effects is a sub model of the adjacent categories location-shift model, which is a sub model of the general model with category-specific effects.

\subsection*{Safety in Naples} 
Figure \ref{tab:nuccatspec2} shows fits for logistic adjacent categories models. It is seen that there is no need to use the general model with category-specific effects since the difference in deviances between the general model and the location-shift version is not significant. However, the location-shift model can not be reduced to the model with global effects. The fits are well comparable to the fits obtained for the cumulative models given in Table \ref{tab:nuccatspec1}. For the adjacent categories as well as for the cumulative modeling approach the location-shift versions turn out to be the best compromise between goodness-of-fit and sparsity. The fits  of the cumulative versions are slightly better than that of the adjacent categories location-shift version.

\begin{table}[!ht]
\caption{Fits of adjacent category models with logistic link for safety data.}\label{tab:nuccatspec2} 
\centering
\begin{tabularsmall}{lrrrrrrrrrrr}
  \toprule
 & deviance     & df  &    deviance & df &$p$-value\\ 
 &      &   &     & \\
  \midrule

Model with category-specific effects               &9828.07     &19935 \\
Location-shift model                      &9902.43     &19998  & 74.36  &63 &0.1549\\
Model with global effects                   &9959.00    &20007  &56.57  &9 &0.0000\\ 
\bottomrule  
\end{tabularsmall}
\end{table}

In the following we briefly compare the alternative modelling approaches for the safety data. 
The estimates of the proportional odds  model and the cumulative logistic location-shift model are given in Table \ref{tab:nuc2}. It is seen that the dispersion effects are not negligible. All variables show rather small $p$-values in the dispersion component, which explains the strong difference in deviances between  the location-shift model and the simple model with global effects, which does not account for varying dispersion. It is seen that the simple proportional odds model   yields stronger location effects than the location-shift model, which is a hint that estimates might be biased if dispersion effects are ignored. The same pattern is found for the adjacent categories models (not given).

\begin{figure}[!ht]
\begin{center}
\includegraphics[width=0.48\textwidth]{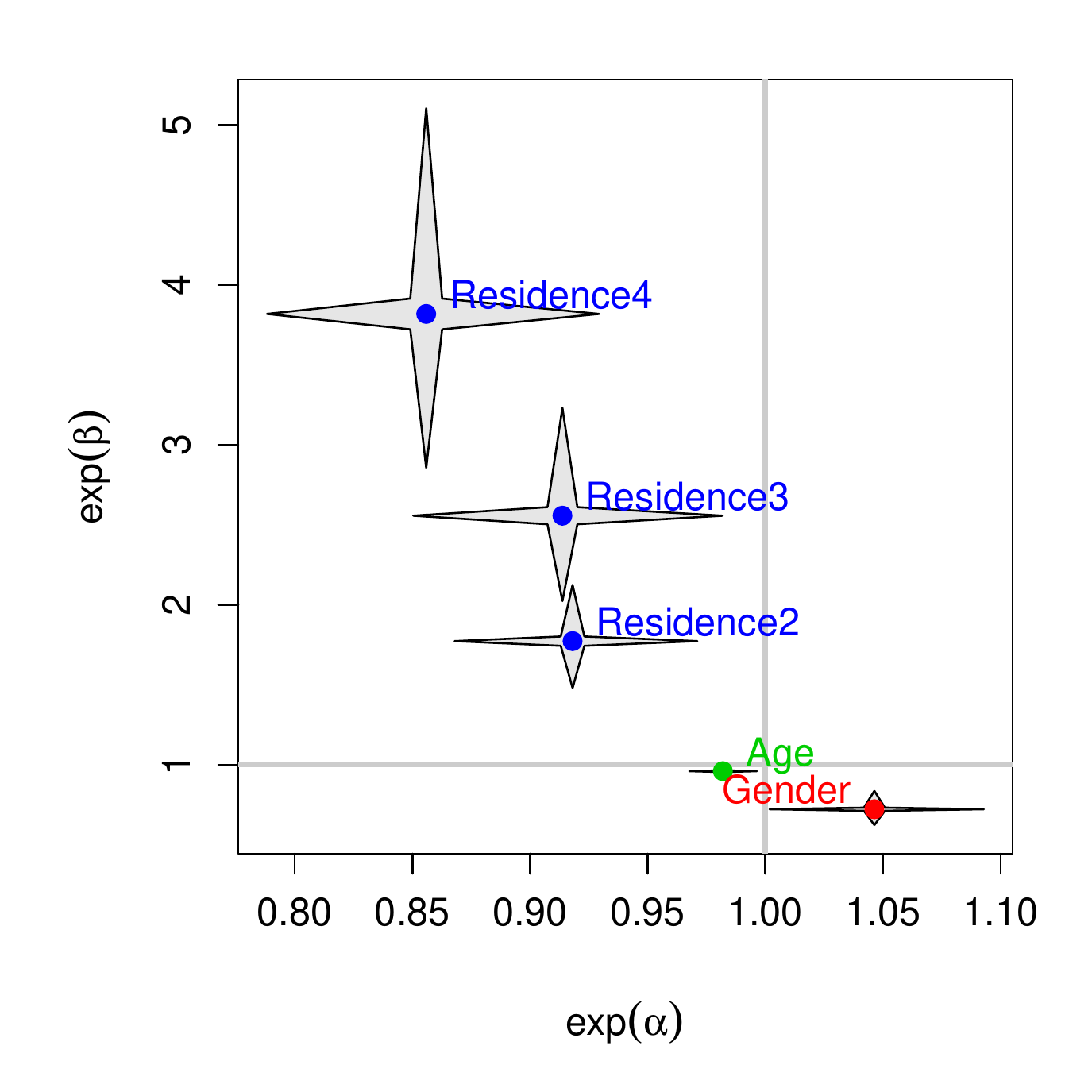}
\includegraphics[width=0.48\textwidth]{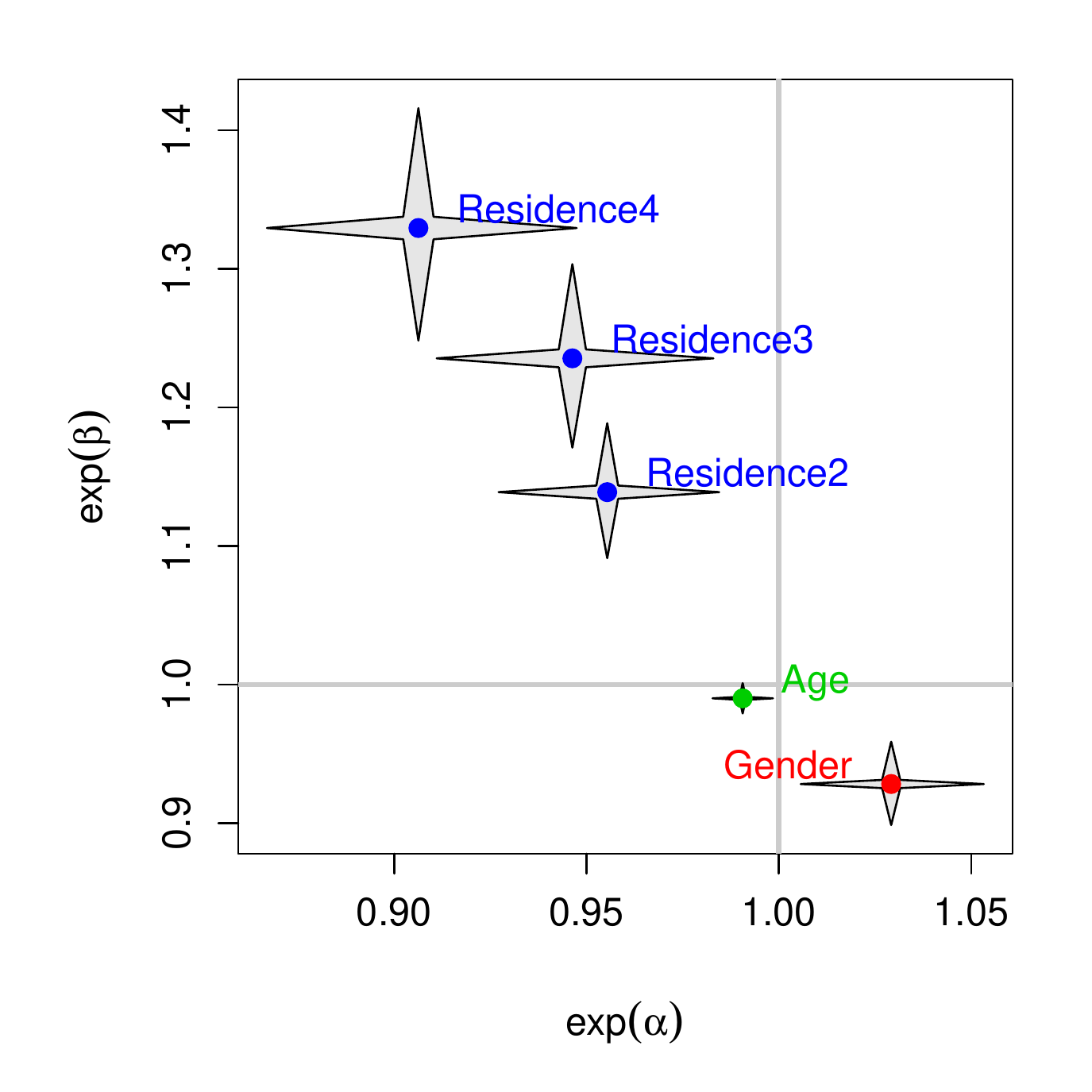}
\caption{Plots of $(e^{\hat{\alpha}},e^{\hat{\beta}})$ for safety data, $y$-axis represents location, $x$-axis represents dispersion, left: cumulative locaion shift model, right: adjacent categories location-shift model.}\label{fig:safetyplott}
\end{center}
\end{figure}

Instead of showing all the parameters we use the plotting tool provided by our R package (see Section \ref{sec:OrdDisp}). In Figure \ref{fig:safetyplott} the location effects of age, gender and residence are plotted against the dispersion effects (left: cumulative model, right: adjacent categories model). The abscissa represents the multiplicative dispersion effect on the odds  $e^{\hat{\alpha}}$, and the ordinate axis represents the multiplicative location effect $e^{\hat{\beta}}$ for the variables. In addition to the point estimates   
pointwise $95\%$ confidence intervals are included. The  horizontal and vertical lengths of the stars correspond  to the confidence intervals of $e^{\hat{\alpha}}$ and $e^{\hat{\beta}}$, respectively. Thus, the stars also show  the significance of effects. If the stars cross the line  $y=1$, location  effects  have to be considered as significant, if they  cross the line $x=1$  dispersion effects have to be considered as significant.

To make the models comparable we did not use the classical representation of the cumulative model. We used the reverse categories representation 
$P(Y \ge r|\xb )= F(\beta_{0r}+\xb^T\betab+ (k/2-r) \zb^T\alphab)$. Then the location effects have the same interpretation as in the adjacent categories model,
large values of $\xb^T\betab$ indicate a preference for high response categories while small values indicate a preference for low categories. 
The resulting star plots for both models the cumulative and the adjacent categories model are very similar. It is seen that people living outside of the city of Naples feel much safer (y axis, reference category: city of Naples), effects are ordered, the larger the distance to the city the safer they feel. Females and older people feel less safe (below the line $y=1$).  As far as  concentration of responses is concerned, people living outside the city of Naples and older people have stronger dispersion (below $x=1$) while females show less dispersion than men. It should be noted that age is measured  in decades, otherwise the age effect would be too close to zero in the star plot. Although the estimated parameter values are quite different for the cumulative and the adjacent categories  models, the conclusions one draws are very similar, as are the goodness-of-fits. Thus one might use either of the two models to investigate the impact of variables. However, it is certainly warranted to account 
for dispersion effects.

\begin{table}[!t]
\caption{ Estimates of proportional odds  model and cumulative logistic location-shift model for  safety data.}\label{tab:nuc2} 
\begin{center}
\begin{tabularsmall}{lrrrrrrrr}
  \toprule
 &\multicolumn{4}{c}{\bf Proportional Odds Model } &\multicolumn{4}{c}{\bf Location-Shift Model }\\
 \midrule
 
 & coef & se & z value & $p$-value & coef & se & z value & $p$-value\\ 
 \midrule
\bf Location effects\\
 \midrule

Age              &-0.045     &0.026   &-1.713   &0.086   &-0.041    &0.026   &-1.578 &0.114   \\
Gender           &-0.343     &0.075   &-4.563 &0.000  &-0.327    &0.075   &-4.343 &0.000 \\
Residence2        &0.518     &0.090    &5.705 &0.000  &0.572    &0.092    &6.199 &0.000  \\
Residence3        &0.899     &0.117    &7.644 &0.000   &0.938    &0.119    &7.859 &0.000  \\
Residence4        &1.397    &0.141    &9.885  &0.000  &1.339    &0.148    &9.039  & 0.000  \\
EduDegree2 &-0.307    &0.111   &-2.748   &0.006  &0.274    &0.112   &-2.449 &0.014   \\
EduDegree3 &-0.319     &0.150   &-2.118   &0.034   &0.294    &0.151   &-1.947 &0.051   \\
EduDegree4 &-0.162     &0.159   &-1.021   &0.307    &0.112    &0.160   &-0.704 &0.481  \\
EduDegree5 &-0.292     &0.221   &-1.319   &0.187   &-0.261    &0.221   &-1.177 &0.239  \\
\midrule
\bf Dispersion effects\\
 \midrule
Age              &&&&&-0.018    &0.007   &-2.442 &0.014 \\
Gender           &&&&&  0.045    &0.022    &2.039 &0.041   \\
Residence2       &&&&&-0.085    &0.028   &-2.984 &0.002  \\
Residence3       &&&&&-0.090    &0.036   &-2.458 &0.013   \\
Residence4       &&&&&-0.155    &0.042   &-3.703 &0.000 \\
EduDegree2  &&&&&0.099    &0.030    &3.297 &0.000 \\
EduDegree3  &&&&&0.142    &0.044    &3.176 &0.001 \\
EduDegree4  &&&&&0.163    &0.047    &3.437 &0.000 \\
EduDegree5  &&&&&0.107    &0.064    &1.668 &0.095 \\
\bottomrule
\end{tabularsmall}
\end{center}
\end{table}

\subsection{Further Applications }
To demonstrate that the location-shift model is frequently a good choice that shows satisfactory goodness-of-fit while being comparably   sparse in parameters we consider some more applications.

\subsubsection*{Nuclear Energy }

The German Longitudinal Election Study (GLES)  is a long-term study of the German electoral process \citep{GLES}.  The data consist of  $2036$  observations and  originate from the pre-election survey for the German federal  election  in  2017 and are concerned with political fears. In particular the participants were   asked: ``How afraid are you due to the use of nuclear energy?''
The answers were measured on Likert scales from 1 (not afraid at all) to 7 (very afraid). The  explanatory variables used in the model are
\textit{Aged} (age of the participant), \textit{Gender} (1: female; 0: male),
\textit{EastWest} (1: East Germany/former GDR; 0: West Germany/former FRG).
Tables \ref{tab:nuccatne1} and \ref{tab:nuccatne2}  show the fits of cumulative and adjacent categories models, respectively. Comparison between the full models with category-specific effects and location-shift versions yields $p$-values greater than 0.8.  It is obvious 
that the location-shift versions of the models represent satisfying approximations while models with global parameters should not be used to describe the underlying response structure. 

\begin{table}[!ht]
\caption{Fits of cumulative models with logistic link for response fear of  nuclear energy.  }\label{tab:nuccatne1} 
\centering
\begin{tabularsmall}{lrrrrrrrrrr}
  \toprule
 & deviance     & df  &  difference in   & df &$p$-value\\ 
 &      &   &deviances     & \\
  \midrule

Model with category-specific effects               &7499.61     &12192 \\
Location-shift model                      &7506.36     &12204  & 6.75  &12 &0.873\\
Model with global effects                   &7544.60    &12206  &38.24  &2 &0.000\\ 
\bottomrule  
\end{tabularsmall}
\end{table}

\begin{table}[!ht]
\caption{Fits of adjacent categories models with logistic link for response fear of  nuclear energy.  }\label{tab:nuccatne2} 
\centering
\begin{tabularsmall}{lrrrrrrrrrr}
  \toprule
 & deviance     & df  &  difference in   & df &$p$-value\\ 
 &      &   &deviances     & \\
  \midrule

Model with category-specific effects               &7500.77     &12192 \\
Location-shift model                      &7508.72     &12204  & 7.95  &12 &0.997\\
Model with global effects                   &7545.41    &12206  &36.69  &2 &0.000\\ 
\bottomrule  
\end{tabularsmall}
\end{table}

\begin{figure}[!ht]
\begin{center}
\includegraphics[width=7cm]{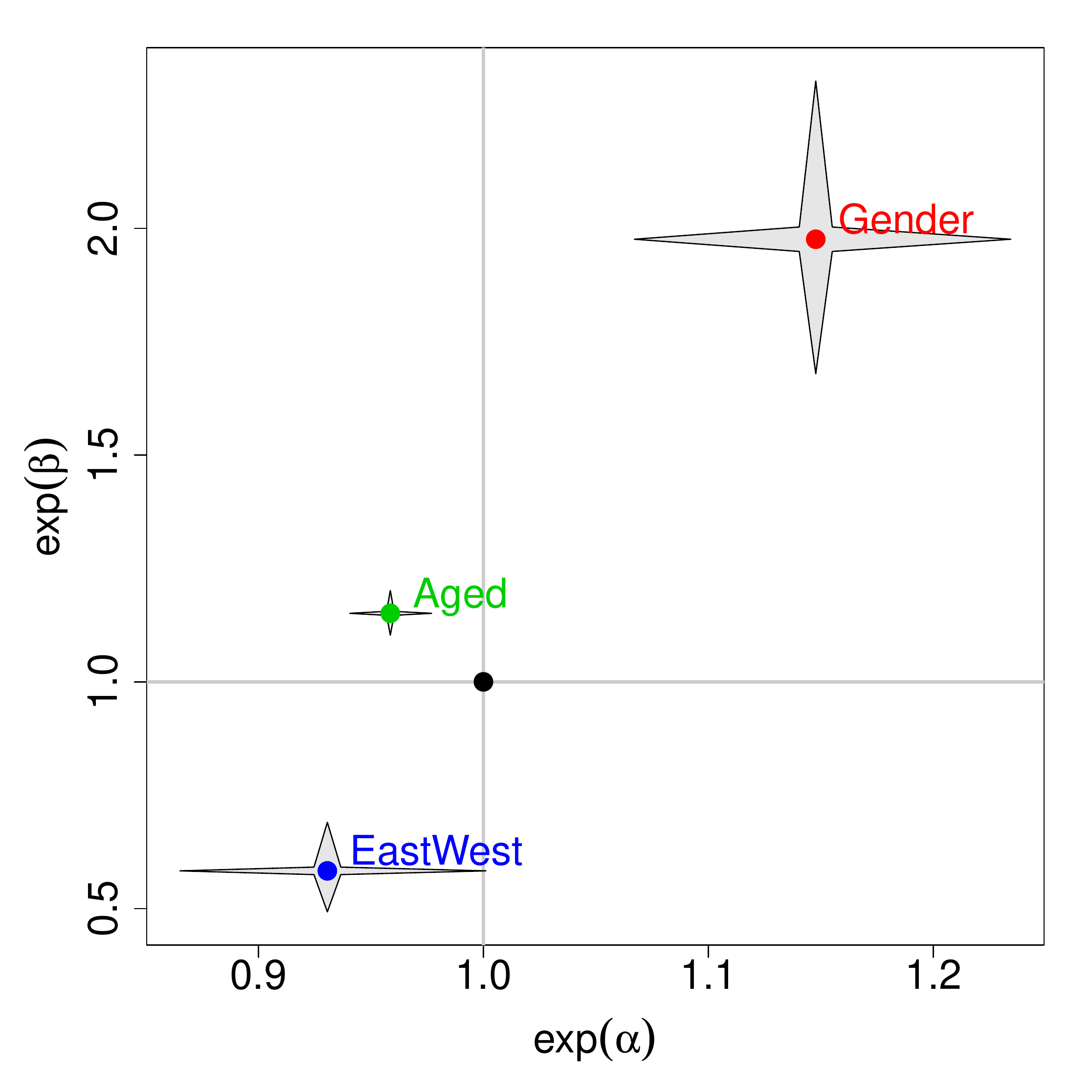}
\includegraphics[width=7cm]{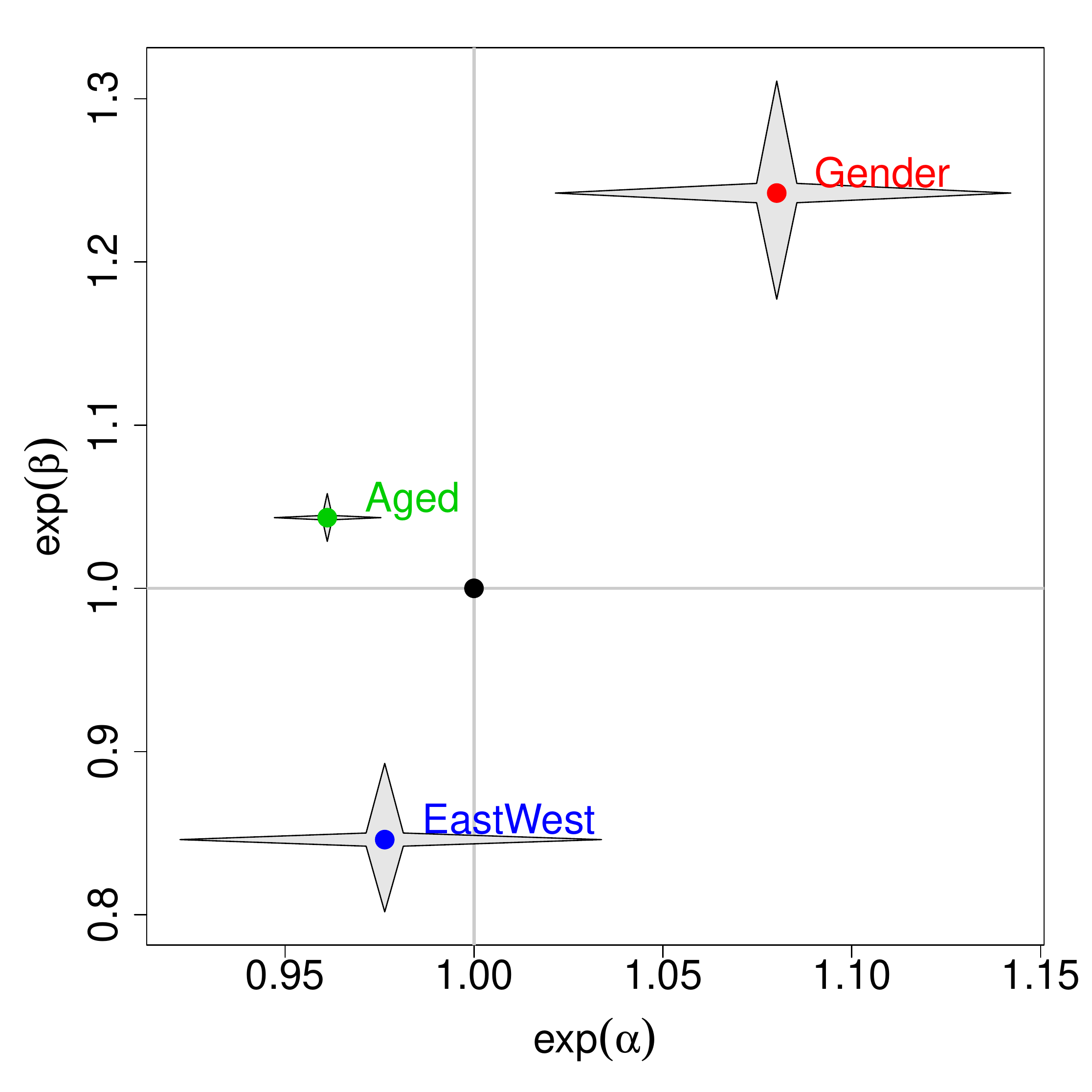}
\caption{Plots of $(e^{\hat{\alpha}},e^{\hat{\beta}})$ for response fear of nuclear energy, $y$-axis represents location, $x$-axis represents dispersion, left: cumulative location shift model, right: adjacent categories location-shift model.}\label{fig:Nuclearplott}
\end{center}
\end{figure}

The strength of effects is seen from the stars in Figure \ref{fig:Nuclearplott}, which shows parameter estimates for the cumulative location-shift model on the left and the adjacent categories model on the right hand side. Again we used the inverse order of categories in the cumulative model and age  measured in decades. It is seen that all parameters have significant location and dispersion effects with the exception of EastWest, for which the dispersion effect is not distinctly significant. It is seen that females and older people are more afraid of the consequences of the use of nuclear energy while residents of the Eastern part are less afraid. Females show stronger dispersion than men, and older people less dispersion than younger respondents.

\subsubsection*{Climate Change }
Let us again consider the GLES data but now the response to the item ``How afraid are you due to the climate change?''.
Tables \ref{tab:nuccats1} and \ref{tab:nuccats2} show the fits of cumulative and adjacent categories models, respectively.
Also for this question  the full models can be simplified to  location-shift versions of the models although the reduction is not so obvious as in the 
question that refers to the use of nuclear energy.

\begin{table}[!ht]
\caption{Fits of cumulative models with logistic link for response fear of  climate change.  }\label{tab:nuccats1} 
\centering
\begin{tabularsmall}{lrrrrrrrrrr}
  \toprule
 & deviance     & df  &  difference in   & df &$p$-value\\ 
 &      &   &deviances     & \\
  \midrule

Model with category-specific effects               &7152.12     &12192 \\
Location-shift model                      &7170.56     &12204  & 18.34  &12 &0.1057\\
Model with global effects                   &7178.06    &12206  &7.50  &2 &0.0235\\ 
\bottomrule  
\end{tabularsmall}
\end{table}

\begin{table}[!ht]
\caption{Fits of adjacent categories models with logistic link for response fear of  climate change.  }\label{tab:nuccats2} 
\centering
\begin{tabularsmall}{lrrrrrrrrrr}
  \toprule
 & deviance     & df  &  difference in   & df &$p$-value\\ 
 &      &   &deviances     & \\
  \midrule

Model with category-specific effects               &7153.42       &12192 \\
Location-shift model                      &7171.93      &12204  & 18.51  &12 &0.1010   \\
Model with global effects                   &7177.17     &12206  &5.24  &2 &0.0723\\ 
\bottomrule  
\end{tabularsmall}
\end{table}

\subsubsection*{Demand for Medical Care}

\citet{DebTri:97} analyzed the demand for medical care for individuals, aged 66 and over, based
on a dataset from the U.S. National Medical Expenditure survey in
1987/88. The data (``NMES1988'') are available from the R package~{AER} \citep{KleZei:2008}. The response is the 
number of physician/non-physician office and hospital outpatient
visits, which is categorized with categories given by 1: zero , 2: 1- 3, 3: 4-6, 4:7-10, 5:11-20, 6: above 20. The available covariates include \textit{Age},  the self-perceived health status (\textit{Health}; 0: poor, 1: average, 2: excellent),  the number of
chronic conditions (\textit{Numchron}). Since the effects vary across gender, we consider only male, married patients ($n=1388$). 
The data set is interesting because it is one of the applications  in which cumulative models show fitting problems. The model with category-specific effects can not be fitted at all, the cumulative location-shift model yields unstable estimates and no standard errors are available. In contrast, for the adjacent categories model maximum likelihood estimates and standard errors are obtained by regular software. The big advantage of the adjacent categories model over the cumulative model that shows here is that  parameter values are not restricted in the adjacent categories model. Table \ref{tab:demacat} shows the fits for the adjacent categories models. It is again seen that one might use the location-shift model but the simple model with global parameters is not appropriate.

\begin{table}[!ht]
\caption{Fits of adjacent categories models with logistic link for demand for medical care data.}\label{tab:demacat} 
\centering
\begin{tabularsmall}{lrrrrrrrrrr}
  \toprule
 & deviance     & df  &  difference in   & df &$p$-value\\ 
 &      &   &deviances     & \\
  \midrule

Non proportional odds model               &4258.41     &6910 \\
Location-shift model                      &4282.05     &6925  & 23.64  &15 &0.0714\\
Proportional odds model                   &4303.64    &6930  &19.69  &5 &0.0014\\ 
\bottomrule  
\end{tabularsmall}
\end{table}

\section{Generalized Additive Models }\label{sec:additive}

In the following traditional additive models for ordinal responses are considered briefly. Then the additive location-shift model is introduced.

\subsection{Generalized Additive Models for Ordinal Responses}
Parametric models as the non proportional odds model are rather restrictive. They assume a simple linear predictor, which might be very misleading if, for example U-shaped effects are present. A very flexible class of models that avoids these restrictions are generalized additive models, which are 
well developed for continuous and univariate responses, see, for example, \citet{BujHasTib:89}, \citet{FriSil:89}, \citet{HasTib:86a}.

In generalized additive models the linear predictor $\xb^T\betab$ is replaced   by the additive term
\[
f_{(1)}(x_1)+\dots +f_{(p)}(x_p),
\]
where the $f_{(j)}(.)$ are unspecified functions. The unknown functions may be expanded in basis functions \citep{EilMar:96}, smoothing splines \citep{Gu:2002} or thin-plate splines \citep{Wood:2004}, all of them have been used to model binary or continuous responses. 

Ordinal models with additive predictors were considered by \citet{YeeWil:96}, \citet{yee2010vgam, yee2015vector} within the framework of vector generalized additive models.
For ordinal models one has to replace the whole predictor $\eta_r=\beta_{0r}+\xb^T\betab$ by 
\[
\eta_r=\beta_{0r}+f_{(1)}(x_1)+\dots +f_{(p)}(x_p),
\]
which contains a category-specific intercept but fixed smooth variable effects.
The essential trait is that it is assumed that the functions $f_{(j)}(x_p)$ do not vary across categories, they are \textit{global} effects. Thus, if one uses the cumulative approach the models can be considered as additive versions of  proportional odds models with accordingly simple interpretation of effects. If the $j$-th variable increases by one unit from $x_j $ to $x_j+1$ and all other variables remain fixed one obtains
\begin{align*}
 &F^{-1}(P(Y\leq r|x_1,\dots, x_j+1,\dots,x_p))- F^{-1}(P(Y\leq r|x_1,\dots, x_j,\dots,x_p))=\\
&=f_{(j)}(x_j+1)-f_{(j)}(x_j),
\end{align*}
which contains only the function $f_{(j)}(.)$. In the logistic version the inverse distribution function is equivalent to the cumulative log-odds, and one obtains
\[
\log\left(\frac{\gamma_r(x_1,\dots, x_j+1,\dots,x_p)}{\gamma_r(x_1,\dots, x_j,\dots,x_p)}\right)= f_{(j)}(x_j+1)-f_{(j)}(x_j),
\]
where the $\gamma_r(\xb)= {P(Y \le r|\xb)}/{P(Y>r|\xb)}$ are the cumulative odds. After transformation one has 
\[
  \frac{\gamma_r(x_1,\dots, x_j+1,\dots,x_p)}{\gamma_r(x_1,\dots, x_j,\dots,x_p)}=  
    e^{f_{(j)}(x_j+1)-f_{(j)}(x_j)},
\]
which can be interpreted as the change in odds if the $j$-th variable increases by one unit from $x_j$ to $x_j+1$.
If the function is linear, that is, $f_{(j)}(x_j)=x_j\beta_j$, one obtains on the right hand side $e^{\beta_j}$, which is equivalent to (\ref{eq:int}). Then, the effect strength does not depend on the  baseline value  $x_j$. This is different in the general additive case, in which the change depends on the 'starting' value $x_j$, which is increased by one unit. Nevertheless, the effect strength is not affected by the values of the other covariates.
Similar properties hold if one uses the adjacent categories model with additive predictor structure.

\subsection{Additive Location-Shift Models}

The additive ordinal models with global effects considered in the previous section share  some problems with the parametric model with global effects.
Although it is more flexible by allowing for smooth effects it is rather restrictive by assuming that the effects of covariates do not depend on the category.
Consequently, it might show bad goodness-of-fit.  One can extend the model in the same way as linear models by allowing that the smooth functions are category-specific. Then, one postulates
\[
\eta_r=\beta_{0r}+f_{(1),r}(x_1)+\dots +f_{(p),r}(x_p),
\]
where the functions $f_{(j),r}(.)$ depend on $r$. However, for each covariate  one has to fit $k-1$ functions, which can lead to a confusing number of functions if one has, for example, 10 response categories, which is not unusual in questionnaires. Moreover,  the functions are severely restricted since $\eta_r\le \eta_{r+1}$has to hold for all $r$, which is hard to control in estimation. If it is not accounted for resulting estimates might yield negative probabilities.

One can try to restrict the variation of the functions by assuming that they are varying not too strongly across categories, see \citet{Tutz:2003}, but this approach calls for complicated regularization methods, and still one has $k-1$ functions to interpret for each explanatory variable.  

The model proposed here is an additive version of the location-shift model, which avoids the large number of functions but typically fits much better than the simple additive model. 
The \textit{additive location-shift model}  uses the predictor
\[
\eta_r=\beta_{0r}+f_{(1)}(x_1)+\dots +f_{(p)}(x_p)+ (r-k/2) \{ f_{(1)}^{(S)}(z_1)+\dots +f_{(m)}^{(S)}(z_m)\},
\]
where $f_{(1)}^{(S)}(.),\dots,f_{(m)}^{(S)}(.)$ are unspecified dispersion functions.
The predictor contains two types of smooth functions, the ones in the location term $f_{(1)}(x_1)+\dots +f_{(p)}(x_p)$, and the ones in the 
dispersion term $f_{(1)}^{(S)}(z_1)+\dots +f_{(m)}^{(S)}(z_m)$. 

In particular when $\xb$ and $\zb$ are distinct the functions have a simple interpretation. If the $j$-th $x$-variable increases by one unit from $x_j $ to $x_j+1$ and all other variables remain fixed one obtains for the cumulative model the same property as in the simple additive model,
\[
  \frac{\gamma_r(x_1,\dots, x_j+1,\dots,x_p)}{\gamma_r(x_1,\dots, x_j,\dots,x_p)}=  
    e^{f_{(j)}(x_j+1)-f_{(j)}(x_j)},
\]
which means that the functions can be interpreted as change in (cumulative) odds ratios.
For the differences between  adjacent predictors one obtains
\[
\eta_r-\eta_{r-1}=\beta_{0r}-\beta_{0,r-1}+  \{f_{(1)}^{(S)}(z_1)+\dots +f_{(m)}^{(S)}(z_m)\}.
\]
That means that large values of $f_{(j)}^{(S)}(.)$ widen  the distance between adjacent predictors, while small values shrink the distance between adjacent predictors. Therefore large values indicate a tendency to middle categories or smaller dispersion while small values indicate a tendency to extreme categories or strong dispersion. 

\blanco{
fraglich:
If the $j$-th $z$-variable increases by one unit from $z_j $ to $z_j+1$ and all other variables remain fixed one obtains for the differences of adjacent predictors
\begin{align*}
&\eta_r(\xb,z_1,\dots, z_j+1,\dots,z_p)-\eta_{r}(\xb,z_1,\dots, z_j,\dots,z_p)=\\
&(r-k/2) \{ f_{(j)}^{(S)}(z_j+1)-f_{(j)}^{(S)}(z_j)\}=(r-k/2) \{ d_{(j)}^{(S)}(z_j)\},
\end{align*}
where $d_{(j)}^{(S)}(z_j)=f_{(j)}^{(S)}(z_j+1)-f_{(j)}^{(S)}(z_j)$ is the change in function $f_{(j)}(z_j)$ if the $j$-th variable increases by one unit.
}

As the parametric model the additive location-shift model accounts for  dispersion without being too complex.  In the general case $\xb=\zb$ the additive location-shift model contains just two  smooth functions per variable that characterize the effect of explanatory variables on the response, one for the location and one for the response style. That means one has to fit only $2p$ smooth functions instead of $(k-1)p$, which would be needed in the general model with category-specific covariate functions. 

For the fitting of the unknown functions $f_{(j)}(.), f_{(j)}^{(S)}(.)$ we use an expansion in basis functions. Thus functions are approximated by
\[
 f_{(j)}(x)= \sum_{s=1}^{M} \beta_{js}\Phi_s(x) \quad f_{(j)}^{(S)}(z) = \sum_{s=1}^{M} \alpha_{js}\Phi_s(z),
\]
where $\Phi_1(.),\dots,\Phi_M(.)$ are basis functions. A widely used set of basis functions are B-splines \citep{EilMar:96}, which are also implemented in our R package \textbf{ordDisp} to be described in detail in Section \ref{sec:OrdDisp}.

\subsection{Safety Data}
\begin{figure}[!ht]
\begin{center}
\includegraphics[width=13cm]{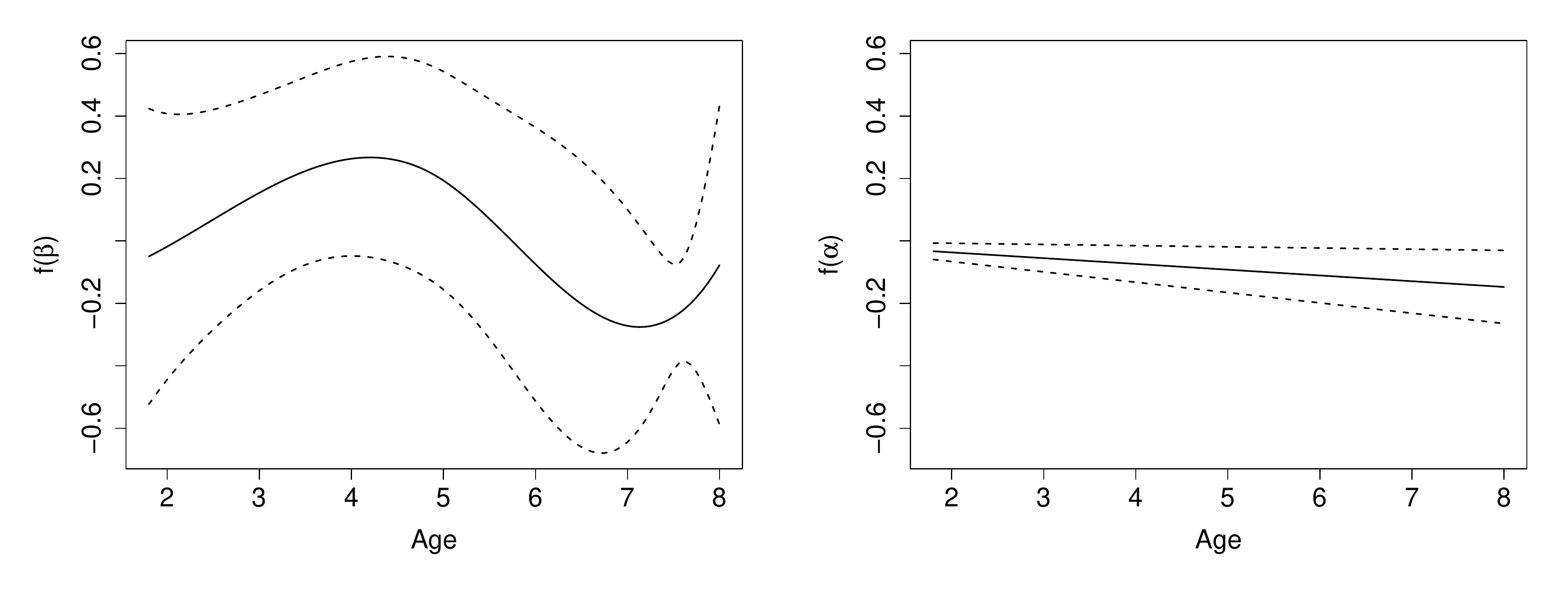}
\includegraphics[width=13cm]{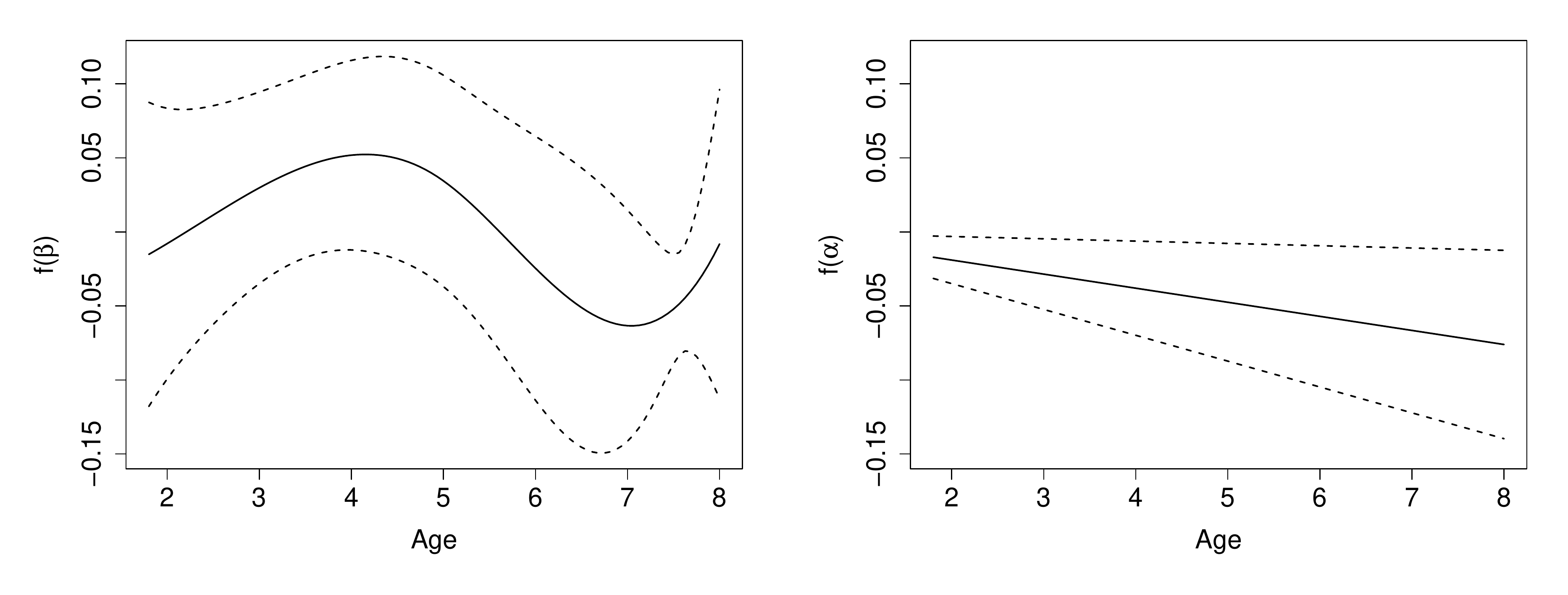}
\caption{Safety data (left: location effect $f_{(age)}(age)$, right: dispersion effect $f_{(age)}^{(S)}(age)$, upper panel: cumulative model, lower panel: adjacent categories model).}\label{fig:safetyplots}
\end{center}
\end{figure}

It has been shown that the parametric location-shift model provided a good compromise between sparsity  and goodness-of-fit for the response feeling safe in Naples.
The only continuous variable was age, which had $p$-values 0.086 (cumulative model) and 0.114  (adjacent categories model). The $p$-values are greater than 0.05 but not so far away that one can be sure that there is no effect of age. In the following age is included as a smooth function approximated by four   cubic  B-splines  in the location term and as a linear function in the dispersion term (a linear function turned out as a good approximation in the dispersion term).  Figure \ref{fig:safetyplots} shows the resulting curves (left: location effect $f_{(age)}(age)$, right: dispersion effect $f_{(age)}^{(S)}(age)$, upper panel: cumulative model, lower panel: adjacent categories model). It is seen that in both models a linear effect of age seems not appropriate. In particular  young and   
older persons seem to feel less safe than persons in their forties. Testing if the smooth effect of age is needed yields a $p$-value of 0.046 (cumulative model), which indicates that age should not be neglected.

\subsection{Nuclear Energy} 
\begin{figure}[!ht]
\begin{center}
\includegraphics[width=12cm]{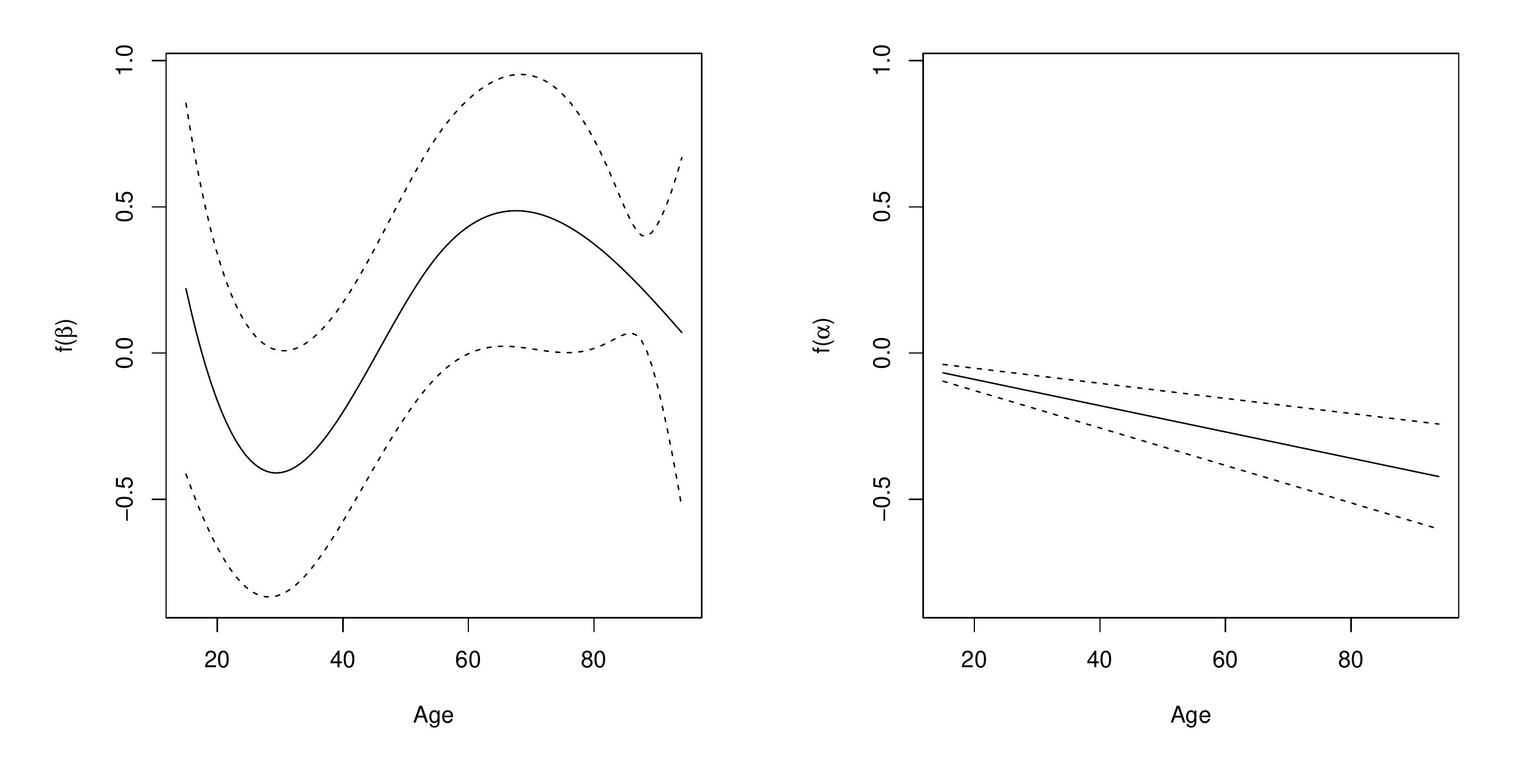}
\includegraphics[width=12cm]{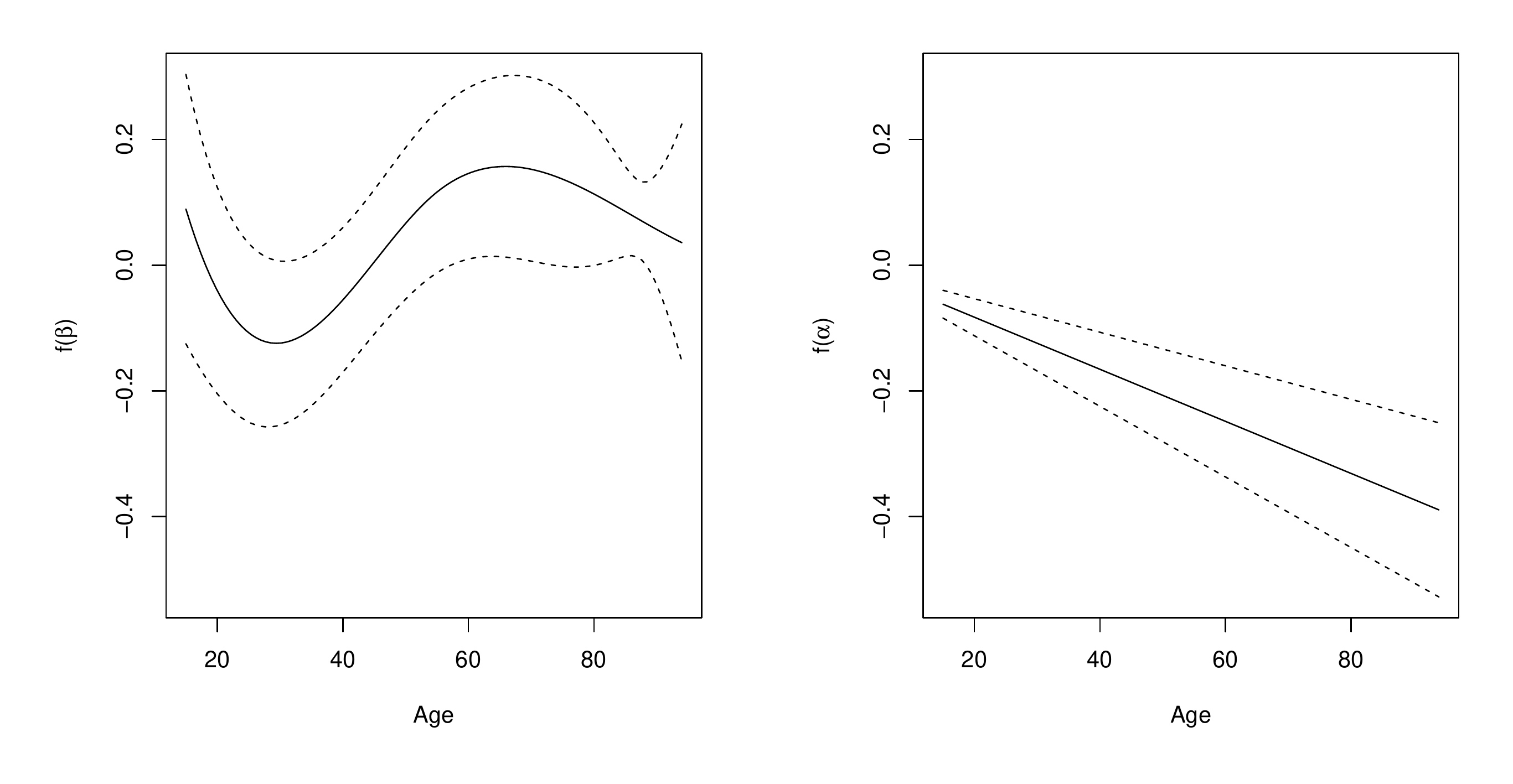}
\caption{Nuclear energy data (left: location effect $f_{(age)}(age)$, right: dispersion effect $f_{(age)}^{(S)}(age)$, upper panel: cumulative model, lower panel: adjacent categories model).}\label{fig:Nuclearplots4}
\end{center}
\end{figure}

As a second example the effect of age on the  fear of the use of nuclear energy is considered. Figure \ref{fig:Nuclearplots4} shows the estimated location and dispersion effects for the additive cumulative and adjacent categories model. The estimates of the location effect indicate that the fear of the use of nuclear energy is strongest for people in their sixties and weakest for people around thirty. The estimates of the dispersion term indicate that older people tend to have less dispersion than younger respondents. Likelihood ratio tests show that location as well as dispersion effects are not to be neglected. The likelihood ratio test for the location effects is 64.62 on 4 df and for the dispersion effect 21.24 on 1 df if the cumulative model is fitted. Similar values result for the adjacent categories model.

\section{Program Package OrdDisp}\label{sec:OrdDisp}

Parametric and additive location-shift models can be fitted using the R add-on package \textbf{ordDisp}~\citep{ordDisp}. The call of the main fitting function (in parts) is the following: 
\begin{verbatim}
ordDisp(formula, data, family = c("cumulative", "acat"), n_bs = 6, 
        reverse = FALSE, ...). 
\end{verbatim}
The formula needs to have the form \texttt{y $\sim$ x1 + ... + xp | z1 + ... + zq}, where on the right hand side of the formula the $x$-variables of the location term and the $z$-variables of the dispersion term are separated by the \texttt{|}-operator. The function allows to fit smooth effects $f(.)$ and $f^{(S)}(.)$ by using \texttt{s(x)} and \texttt{s(z)} in the respective part of the formula. The functions are then fitted using \texttt{n\_bs} B-spline basis functions. In the case of nominal covariates \texttt{ordDisp()} generates 0-1-coded dummy variables. If \texttt{reverse=TRUE} the function uses the reverse categories representation $P(Y \ge r|\xb)/P(Y < r|\xb)$ for the cumulative model and $P(Y = r|\xb)/P(Y = r+1|\xb)$ for the adjacent categories model. To keep the interpretation of the dispersion effects the scaling factor is reversed to $(k/2-r)$ in the cumulative case and $(r-k/2)$ in the adjacent categories case.

Function \texttt{ordDisp()} internally calls function \texttt{vglm()} of the R package \textbf{VGAM}~\citep{yee2010vgam}. Thus the fitted object inherits all the values of a \texttt{vglm}-object and importantly all the methods implemented for objects of class \texttt{vglm}, like \texttt{print}, \texttt{summary}, \texttt{predict} and \texttt{plot} can be applied. Additionally, star plots depicting the location effects against the dispersion effects including pointwise 95\% confidence intervals (cf. Figure \ref{fig:safetyplott}) can be generated using the function \texttt{plotordDisp(object, names, ...)}, where the variables to be plotted are passed to the function by the \texttt{names}-argument. Note that the use of \texttt{plotordDisp()} is only meaningful for variables with both a location effect and a dispersion effect.  

\section{Concluding Remarks}\label{sec:conclud}
It has been demonstrated that parametric location-shift models typically are sufficiently complex to approximate the underlying probability structure in ordinal regression. The models were also extended to allow for a more general additive predictor structure that avoids to fit a large number of functions. 

Implicitly we compared two approaches to modeling data, the cumulative approach and the adjacent categories approach. Typically the models yield similar goodness-of-fit. A distinct advantage of the adjacent categories model is that no restrictions on the parameter space are postulated, which makes it more adequate when allowing for a more complex predictor structure. 

There is a third class of ordinal models, namely sequential models, which have not been considered here. Parametric sequential models have the form $P(Y \ge r|Y \ge r-1, \xb)= F(\beta_{0r}+\xb^T\betab_r)$. They  reflect the successive transition to higher categories in a stepwise fashion since $Y \ge r$  given $Y \ge r-1$ can be interpreted as the transition  to categories higher than category $r-1$ given at least category $r-1$  
has been reached. Sequential models are strongly linked to discrete survival, and have been considered, for example, by \citet{ArmSlo:89}, \citet{Tutz:91c},  \citet{AnaKlei:97}. Location-shift models for this type of model seem less useful because of the structure of the model. In sequential models the category-specific parameters $\betab_r$ have a distinct meaning, they represent the impact of covariates on the transition to   higher categories given lower categories have already been reached. Including a shift term, which represents a tendency to middle or extreme categories seems less useful.

\bibliography{literatur}

\begin{thebibliography}{}

\bibitem[\protect\citeauthoryear{Ananth and Kleinbaum}{Ananth and
  Kleinbaum}{1997}]{AnaKlei:97}
Ananth, C.~V. and D.~G. Kleinbaum (1997).
\newblock Regression models for ordinal responses: A review of methods and
  applications.
\newblock {\em International Journal of Epidemiology\/}~{\em 26}, 1323--1333.

\bibitem[\protect\citeauthoryear{Armstrong and Sloan}{Armstrong and
  Sloan}{1989}]{ArmSlo:89}
Armstrong, B. and M.~Sloan (1989).
\newblock Ordinal regression models for epidemiologic data.
\newblock {\em American Journal of Epidemiology\/}~{\em 129}, 191--204.

\bibitem[\protect\citeauthoryear{Bender and Grouven}{Bender and
  Grouven}{1998}]{BenGro:98}
Bender, R. and U.~Grouven (1998).
\newblock Using binary logistic regression models for ordinal data with
  non--proportional odds.
\newblock {\em Journal of Clinical Epidemiology\/}~{\em 51}, 809--816.

\bibitem[\protect\citeauthoryear{Berger}{Berger}{2020}]{ordDisp}
Berger, M. (2020).
\newblock {\em ordDisp: Separating Location and Dispersion in Ordinal
  Regression Models}.
\newblock R package version 2.1.1.

\bibitem[\protect\citeauthoryear{Brant}{Brant}{1990}]{Brant:90}
Brant, R. (1990).
\newblock Assessing proportionality in the proportional odds model for ordinal
  logistic regression.
\newblock {\em Biometrics\/}~{\em 46}, 1171--1178.

\bibitem[\protect\citeauthoryear{Buja, Hastie, and Tibshirani}{Buja
  et~al.}{1989}]{BujHasTib:89}
Buja, A., T.~Hastie, and R.~Tibshirani (1989).
\newblock Linear smoothers and additive models.
\newblock {\em Annals of Statistics\/}~{\em 17}, 453--510.

\bibitem[\protect\citeauthoryear{Campbell and Donner}{Campbell and
  Donner}{1989}]{CamDon:89}
Campbell, M.~K. and A.~P. Donner (1989).
\newblock Classification efficiency of multinomial logistic-regression relative
  to ordinal logistic-regression.
\newblock {\em Journal of the American Statistical Association\/}~{\em
  84\/}(406), 587--591.

\bibitem[\protect\citeauthoryear{Cox}{Cox}{1995}]{Cox:95}
Cox, C. (1995).
\newblock Location-scale cumulative odds models for ordinal data: A generalized
  non-linear model approach.
\newblock {\em Statistics in Medicine\/}~{\em 14}, 1191--1203.

\bibitem[\protect\citeauthoryear{Deb and Trivedi}{Deb and
  Trivedi}{1997}]{DebTri:97}
Deb, P. and P.~K. Trivedi (1997).
\newblock Demand for medical care by the elderly: A finite mixture approach.
\newblock {\em Journal of Applied Econometrics\/}~{\em 12\/}(3), 313--336.

\bibitem[\protect\citeauthoryear{Eilers and Marx}{Eilers and
  Marx}{1996}]{EilMar:96}
Eilers, P. H.~C. and B.~D. Marx (1996).
\newblock Flexible smoothing with {B}-splines and {P}enalties.
\newblock {\em Statistical Science\/}~{\em 11}, 89--121.

\bibitem[\protect\citeauthoryear{Friedman and Silverman}{Friedman and
  Silverman}{1989}]{FriSil:89}
Friedman, J.~H. and B.~Silverman (1989).
\newblock Flexible parsimonious smoothing and additive modelling (with
  discussion).
\newblock {\em Technometrics\/}~{\em 31}, 3--39.

\bibitem[\protect\citeauthoryear{Genter and Farewell}{Genter and
  Farewell}{1985}]{GenFar:85}
Genter, F.~C. and V.~T. Farewell (1985).
\newblock Goodness-of-link testing in ordinal regression models.
\newblock {\em Canadian Journal of Statistics\/}~{\em 13}, 37--44.

\bibitem[\protect\citeauthoryear{Gu}{Gu}{2002}]{Gu:2002}
Gu, C. (2002).
\newblock {\em Smoothing Splines ANOVA Models}.
\newblock New York: Springer--Verlag.

\bibitem[\protect\citeauthoryear{Hastie and Tibshirani}{Hastie and
  Tibshirani}{1986}]{HasTib:86a}
Hastie, T. and R.~Tibshirani (1986).
\newblock Generalized additive models (c/r: p.~310--318).
\newblock {\em Statist. Sci.\/}~{\em 1}, 297--310.

\bibitem[\protect\citeauthoryear{Iannario, Monti, Piccolo, Ronchetti,
  et~al.}{Iannario et~al.}{2017}]{iannario2017robust}
Iannario, M., A.~C. Monti, D.~Piccolo, E.~Ronchetti, et~al. (2017).
\newblock Robust inference for ordinal response models.
\newblock {\em Electronic Journal of Statistics\/}~{\em 11\/}(2), 3407--3445.

\bibitem[\protect\citeauthoryear{Iannario, Piccolo, and Simone}{Iannario
  et~al.}{2015}]{iannario2018cub}
Iannario, M., D.~Piccolo, and R.~Simone (2015).
\newblock {CUB}: a class of mixture models for ordinal data. r package version
  1.1.3, http://cran.r-project.org/package=cub.

\bibitem[\protect\citeauthoryear{Kim}{Kim}{2003}]{kim2003assessing}
Kim, J.-H. (2003).
\newblock Assessing practical significance of the proportional odds assumption.
\newblock {\em Statistics \& probability letters\/}~{\em 65\/}(3), 233--239.

\bibitem[\protect\citeauthoryear{Kleiber and Zeileis}{Kleiber and
  Zeileis}{2008}]{KleZei:2008}
Kleiber, C. and A.~Zeileis (2008).
\newblock {\em Applied Econometrics with R}.
\newblock New York: Springer--Verlag.

\bibitem[\protect\citeauthoryear{Liu, Mukherjee, Suesse, Sparrow, and Park}{Liu
  et~al.}{2009}]{liu2009graphical}
Liu, I., B.~Mukherjee, T.~Suesse, D.~Sparrow, and S.~K. Park (2009).
\newblock Graphical diagnostics to check model misspecification for the
  proportional odds regression model.
\newblock {\em Statistics in medicine\/}~{\em 28\/}(3), 412--429.

\bibitem[\protect\citeauthoryear{McCullagh}{McCullagh}{1980}]{McCullagh:80}
McCullagh, P. (1980).
\newblock Regression model for ordinal data (with discussion).
\newblock {\em Journal of the Royal Statistical Society\/}~{\em B 42},
  109--127.

\bibitem[\protect\citeauthoryear{Peterson and Harrell}{Peterson and
  Harrell}{1990}]{PetHar:90}
Peterson, B. and F.~E. Harrell (1990).
\newblock Partial proportional odds models for ordinal response variables.
\newblock {\em Applied Statistics\/}~{\em 39}, 205--217.

\bibitem[\protect\citeauthoryear{Rattinger, Ro{\ss}teutscher, Schmitt-Beck,
  We{\ss}els, and Wolf}{Rattinger et~al.}{2014}]{GLES}
Rattinger, H., S.~Ro{\ss}teutscher, R.~Schmitt-Beck, B.~We{\ss}els, and C.~Wolf
  (2014).
\newblock Pre-election cross section {(GLES 2013)}.
\newblock {\em GESIS Data Archive, Cologne\/}~{\em ZA5700 Data file Version
  2.0.0}.

\bibitem[\protect\citeauthoryear{Rudolfer, Watson, and Lesaffre}{Rudolfer
  et~al.}{1995}]{Rud-etal:95}
Rudolfer, S.~M., P.~C. Watson, and E.~Lesaffre (1995).
\newblock Are ordinal models useful for classification? a revised analysis.
\newblock {\em Journal of Statistical Computation Simulation\/}~{\em 52\/}(2),
  105--132.

\bibitem[\protect\citeauthoryear{Snell}{Snell}{1964}]{Snell:64}
Snell, E.~J. (1964).
\newblock A scaling procedure for ordered categorical data.
\newblock {\em Biometrics\/}~{\em 20}, 592--607.

\bibitem[\protect\citeauthoryear{Steadman and Weissfeld}{Steadman and
  Weissfeld}{1998}]{SteWei:98}
Steadman, S. and L.~Weissfeld (1998).
\newblock A study of the effect of dichotomizing ordinal data upon modelling.
\newblock {\em Communications in Statistics -- Simulation and
  Computation\/}~{\em 27(4)}, 871--887.

\bibitem[\protect\citeauthoryear{Tutz}{Tutz}{1991}]{Tutz:91c}
Tutz, G. (1991).
\newblock Sequential models in ordinal regression.
\newblock {\em Computational Statistics \& Data Analysis\/}~{\em 11}, 275--295.

\bibitem[\protect\citeauthoryear{Tutz}{Tutz}{2003}]{Tutz:2003}
Tutz, G. ({2003}).
\newblock {Generalized semiparametrically structured ordinal models}.
\newblock {\em {Biometrics}\/}~{\em {59}}, {263--273}.

\bibitem[\protect\citeauthoryear{Tutz}{Tutz}{2012}]{TutzBook2011}
Tutz, G. (2012).
\newblock {\em {Regression for Categorical Data}}.
\newblock Cambridge University Press.

\bibitem[\protect\citeauthoryear{Tutz and Berger}{Tutz and
  Berger}{2016}]{TuBerg2016RespStyle}
Tutz, G. and M.~Berger (2016).
\newblock Response styles in rating scales - simultaneous modelling of
  content-related effects and the tendency to middle or extreme categories.
\newblock {\em Journal of Educational and Behavioral Statistics\/}~{\em 41},
  239--268.

\bibitem[\protect\citeauthoryear{Tutz and Berger}{Tutz and
  Berger}{2017}]{TuBer2017Disp}
Tutz, G. and M.~Berger (2017).
\newblock Separating location and dispersion in ordinal regression models.
\newblock {\em Econometrics and Statistics\/}~{\em 2}, 131--148.

\bibitem[\protect\citeauthoryear{Walker}{Walker}{2016}]{walker2016generalizing}
Walker, R.~W. (2016).
\newblock On generalizing cumulative ordered regression models.
\newblock {\em Journal of Modern Applied Statistical Methods\/}~{\em 15\/}(2),
  28.

\bibitem[\protect\citeauthoryear{Walker and Duncan}{Walker and
  Duncan}{1967}]{WalDun:67}
Walker, S.~H. and D.~B. Duncan (1967).
\newblock Estimation of the probability of an event as a function of several
  independent variables.
\newblock {\em Biometrika\/}~{\em 54}, 167--178.

\bibitem[\protect\citeauthoryear{Williams and Grizzle}{Williams and
  Grizzle}{1972}]{WilGri:72}
Williams, O.~D. and J.~E. Grizzle (1972).
\newblock Analysis of contingency tables having ordered response categories.
\newblock {\em Journal of the American Statistical Association\/}~{\em 67},
  55--63.

\bibitem[\protect\citeauthoryear{Williams}{Williams}{2006}]{williams2006generalized}
Williams, R. (2006).
\newblock Generalized ordered logit/partial proportional odds models for
  ordinal dependent variables.
\newblock {\em Stata Journal\/}~{\em 6\/}(1), 58.

\bibitem[\protect\citeauthoryear{Williams}{Williams}{2016}]{williams2016understanding}
Williams, R. (2016).
\newblock Understanding and interpreting generalized ordered logit models.
\newblock {\em The Journal of Mathematical Sociology\/}~{\em 40\/}(1), 7--20.

\bibitem[\protect\citeauthoryear{Wood}{Wood}{2004}]{Wood:2004}
Wood, S.~N. (2004).
\newblock Stable and efficient multiple smoothing parameter estimation for
  generalized additive models.
\newblock {\em Journal of the American Statistical Association\/}~{\em 99},
  673--686.

\bibitem[\protect\citeauthoryear{Yee}{Yee}{2010}]{yee2010vgam}
Yee, T. (2010).
\newblock {The VGAM package for categorical data analysis}.
\newblock {\em Journal of Statistical Software\/}~{\em 32\/}(10), 1--34.

\bibitem[\protect\citeauthoryear{Yee}{Yee}{2015}]{yee2015vector}
Yee, T.~W. (2015).
\newblock {\em Vector generalized linear and additive models: with an
  implementation in R}.
\newblock Springer.

\bibitem[\protect\citeauthoryear{Yee and Wild}{Yee and Wild}{1996}]{YeeWil:96}
Yee, T.~W. and C.~J. Wild (1996).
\newblock Vector generalized additive models.
\newblock {\em Journal of the Royal Statistical Society\/}~{\em B}, 481--493.

\end{thebibliography}
\end{document}